\documentclass{aa}
\pdfoutput=1
\usepackage{txfonts}
\usepackage{url}
\usepackage{hyperref}
\usepackage{natbib}
\usepackage{graphicx}
\usepackage{multicol}
\usepackage{float}

\begin{document}

\defcitealias{hildebrandt/etal:2020}{H20}
\defcitealias{wright/etal:2020}{W20a}

\title{KiDS-1000 catalogue: Redshift distributions and their calibration}

\author
{
H.~Hildebrandt\inst{1}
\and
J.~L.~van~den~Busch\inst{1,2}
\and
A.~H.~Wright\inst{1}
\and
C.~Blake\inst{3}
\and
B.~Joachimi\inst{4}
\and
K.~Kuijken\inst{5}
\and
T.~Tr\"oster\inst{6}
\and
M.~Asgari\inst{6}
\and
M.~Bilicki\inst{7}
\and
J.~T.~A.~de~Jong\inst{8}
\and
A.~Dvornik\inst{1}
\and
T.~Erben\inst{2}
\and
F.~Getman\inst{9}
\and
B.~Giblin\inst{6}
\and
C.~Heymans\inst{6,1}
\and
A.~Kannawadi\inst{10}
\and
C.-A.~Lin\inst{6}
\and
H.-Y.~Shan\inst{11,12}
}

\authorrunning{The KiDS collaboration}

\institute{
Ruhr University Bochum, Faculty of Physics and Astronomy, Astronomical Institute (AIRUB), German Centre for Cosmological Lensing, 44780 Bochum, Germany,\\ \email{hendrik@astro.ruhr-uni-bochum.de}
\and
Argelander-Institut f\"ur Astronomie, Universit\"at Bonn, Auf dem H\"ugel 71, 53121 Bonn, Germany
\and
Centre for Astrophysics \& Supercomputing, Swinburne University of
Technology, PO Box 218, Hawthorn, VIC 3122, Australia
\and
Department of Physics and Astronomy, University College London, Gower Street, London WC1E 6BT, UK
\and
Leiden Observatory, Leiden University, Niels Bohrweg 2, 2333 CA
Leiden, the Netherlands
\and
Institute for Astronomy, University of Edinburgh, Royal Observatory, Blackford Hill, Edinburgh EH9 3HJ, UK
\and
Center for Theoretical Physics, Polish Academy of Sciences, al. Lotnik\'ow 32/46, 02-668, Warsaw, Poland
\and
Kapteyn Astronomical Institute, University of Groningen, 9700AD
Groningen, the Netherlands
\and
INAF - Astronomical Observatory of Capodimonte, Via Moiariello 16, 80131 Napoli, Italy	
\and
Department of Astrophysical Sciences, Princeton University, 4 Ivy Lane, Princeton, NJ 08544, USA
\and
Shanghai Astronomical Observatory (SHAO), Nandan Road 80, Shanghai
200030, China
\and
University of Chinese Academy of Sciences, Beijing 100049, China
}

\date{Received 2020-07-30; accepted 2021-01-20}

\abstract{We present redshift distribution estimates of galaxies
  selected from the fourth data release of the Kilo-Degree Survey over
  an area of $\sim1000$\,deg$^2$ (KiDS-1000). These redshift distributions
  represent one of the crucial ingredients for weak gravitational
  lensing measurements with the KiDS-1000 data. The primary estimate
  is based on deep spectroscopic reference catalogues that are
  re-weighted with the help of a self-organising map (SOM) to closely
  resemble the KiDS-1000 sources, split into five tomographic redshift
  bins in the photometric redshift range
  $0.1<z_\mathrm{B}\le1.2$. Sources are selected such that they only
  occupy that volume of nine-dimensional magnitude-space that is
  also covered by the reference samples (`gold' selection). Residual
  biases in the mean redshifts determined from this calibration are
  estimated from mock catalogues to be $\lesssim0.01$ for all five
  bins with uncertainties of $\sim 0.01$. This primary SOM estimate of
  the KiDS-1000 redshift distributions is complemented with an
  independent clustering redshift approach. After validation of the
  clustering-$z$ on the same mock catalogues and a careful assessment
  of systematic errors, we find no significant bias of the SOM
  redshift distributions with respect to the clustering-$z$
  measurements. The SOM redshift distributions re-calibrated by the
  clustering-$z$ represent an alternative calibration of the redshift
  distributions with only slightly larger uncertainties in the mean
  redshifts of $\sim 0.01-0.02$ to be used in KiDS-1000 cosmological
  weak lensing analyses. As this includes the SOM uncertainty,
  clustering-$z$ are shown to be fully competitive on KiDS-1000 data.}

\maketitle

\section{Introduction}

One of the most important goals of observational astronomy has always been to add a third dimension to the two-dimensional images of the sky. In modern extra-galactic imaging surveys containing tens of millions of galaxies this information is obtained with the technique of photometric redshifts \citep[photo-$z$; see][for a recent review]{salvato/etal:2019}. The cosmological redshift leads to a reddening of galaxy spectra that can be detected by observing a galaxy in different photometric pass-bands. This can yield approximate redshifts at a much higher efficiency and to fainter magnitudes than any spectroscopic technique, albeit at significantly reduced precision.

Measurements of weak gravitational lensing \citep[WL; see e.g.][]{bartelmann/etal:2001} crucially depend on photo-$z$ to estimate the geometric factors that enter the modelling of this effect. The accuracy of the cosmological conclusions drawn from modern WL surveys depends directly on the accuracy of the photo-$z$ used to model the WL observables \citep{huterer/etal:2006}. In this process, it is useful to distinguish two different regimes where multi-band photometric information is typically used \citep{newman/etal:2015}. First, approximate individual redshifts for all galaxies used in a WL measurement are estimated to bin the galaxies along the redshift axis. Secondly, the same photometric information is used to estimate the redshift distributions of the ensembles of galaxies in these so-called tomographic bins. While high precision is desirable for the first task in order to attain a high resolution along the line-of-sight, accuracy of the second task determines the quality of the cosmological estimates from statistical WL measurements in the end.

Recently, most surveys have employed template-based techniques (that assume a physical model) to tackle the first problem and empirical or machine-learning (ML) techniques (using e.g. a spectroscopic calibration sample) for the second problem. These choices follow directly from the requirements for the individual photo-$z$ (low scatter) and the redshift distributions of ensembles of galaxies (low bias).

In this paper, we concentrate on the second problem and how it is solved for the cosmological analysis of the KiDS-1000 data set based on the fourth data release \citep{kuijken/etal:2019} of the Kilo-Degree Survey \citep[KiDS;][]{dejong/etal:2013}. The two other ongoing stage-III surveys are the Dark Energy Survey \citep[DES;][]{flaugher/etal:2015} and the Hyper Suprime-Cam Subaru Strategic Program \citep[HSC][]{aihara/etal:2017} survey, whose redshift calibrations in their most recent cosmological analyses are described in \citet{hoyle/etal:2018} and \citet{tanaka/etal:2018}, respectively.

The main requirement is to get an unbiased estimate of the mean redshift of the galaxies in the different tomographic bins \citep[e.g.][]{laureijs/etal:2011}. The uncertainty on this mean redshift needs to be of the order of $\sigma_{\langle z \rangle}\sim 0.01$ for stage-III surveys to not seriously jeopardise their constraining power \citep[see Appendix A of][]{hildebrandt/etal:2017}. These uncertainties are propagated into the full error budget and the exact requirement depends on which survey is analysed and which degradation with respect to the pure statistical uncertainty is deemed acceptable.

Certainly, higher-order moments of the redshift distributions also play a role. However, as WL is an integrated effect along the line-of-sight, the accuracy in estimating these higher-order moments is less important than the mean redshift and can, under normal conditions, be ignored for stage-III surveys \citep[see e.g.][]{hoyle/etal:2018}; it will become important though for upcoming stage-IV experiments like the ESA/NASA \emph{Euclid} space mission \citep{laureijs/etal:2011}, the Vera C. Rubin Observatory Legacy Survey of Space and Time \citep[LSST;][]{ivezic/etal:2019}, and the Nancy Grace Roman Space Telescope \citep[RST;][]{spergel/etal:2015}. For these future missions, not only do the mean redshifts need to be controlled to $\sigma_{\langle z \rangle}\sim 0.001-0.002$, but the shape of the distribution also needs to be known accurately. At this level of precision, it also becomes relevant that the redshift distributions vary spatially with observing conditions so that even redshift distributions that perfectly describe the average of a survey are no longer sufficient and additional corrections are needed \citep{heydenreich/etal:2020}. Similarly, correlations between the point spread function (PSF) ellipticity and the accuracy of the redshift measurements can no longer be ignored \citep{asgari/etal:2019}.

The calibration of the redshifts is usually achieved with the help of a reference sample, which is often spectroscopic but can occasionally also utilise higher-quality photo-$z$, like the COSMOS-2015 catalogue \citep{laigle/etal:2016} based on photometry from more than 30 bands. These reference samples are selected in different ways than the WL source samples and hence are in general neither complete nor representative of those source samples. Different techniques have been developed to overcome this limitation, mostly through re-weighting the reference samples and through culling of the source samples. \citet{lima/etal:2008} describe a re-weighting approach that utilises a $k$-nearest neighbour search in multi-dimensional magnitude-space to re-weight a spectroscopic sample such that it resembles a target photometric sample with unknown redshifts, which are the WL sources in our case. This approach was tested in \citet{cunha/etal:2009} and later on used for KiDS in \citet{hildebrandt/etal:2017,hildebrandt/etal:2020} as well as for DES \citep{bonnett/etal:2016,hoyle/etal:2018}, and to some degree also for HSC \citep{tanaka/etal:2018}. The estimated uncertainties with this approach are of the order of $\sigma_{\langle z \rangle}\sim 0.02$, which was sufficient for the first cosmological analyses that used only a fraction of the data from the stage-III surveys. In order to fully exploit the statistical power of the completed surveys, these uncertainties have to be improved by a factor of $\sim2$.

A similar re-weighting can be achieved by projecting the multi-dimensional magnitude-space into two dimensions with the help of a self-organising map \citep[SOM; ][]{kohonen:1982}. This was pioneered in the framework of Euclid by \citet{masters/etal:2015} and is now being used by KiDS \citep[][hereafter W20a]{wright/etal:2020} and HSC \citep{tanaka/etal:2018}, and suggested for DES \citep{buchs/etal:2019}, too. These studies show that the SOM can, under certain conditions, reach uncertainties in the mean redshifts of tomographic bins of $\sigma_{\langle z \rangle}\sim 0.01$, or even better. Thus, it represents a very promising technique to calibrate redshifts for WL applications in current and future projects.

A complementary estimate of the source redshift distribution can be obtained through cross-correlation studies \citep{schneider/etal:2006,newman:2008,matthews/etal:2010,schmidt/etal:2013,menard/etal:2013,mcquinn/etal:2013,morrison/etal:2017,johnson/etal:2017,davis/etal:2017,scottez/etal:2018,gatti/etal:2018}. Here, the colour-information is not used but instead the angular cross-correlation of the positions of a target source sample and a reference sample with known redshifts is employed. The appeal of this technique is that, unlike colour-based methods, the reference and target samples need not share any magnitude-space whatsoever. All galaxies at a given redshift cluster with each other and hence, in principle a bright reference sample that is relatively easy to observe spectroscopically can be used to calibrate the redshift distribution of a faint source sample. Besides spatial overlap on the sky, the most important requirement is that the reference sample covers the whole redshift range that needs to be probed for the target sample. An important nuisance in this method is the presence of galaxy bias: The fact that galaxies are biased tracers of the underlying matter field can influence the measured cross-correlation functions in a systematic fashion. For the purpose of estimating the redshift distribution, the absolute value of the galaxy bias can usually be neglected (its effect is removed through normalisation of the redshift distribution), any redshift evolution of the galaxy bias must be corrected \citep[e.g.][]{newman:2008,schmidt/etal:2013}.

Clustering-redshift (clustering-$z$) measurements in the literature differ in the details of the implementation of the measurement itself as well as the galaxy bias correction scheme. All these approaches have one thing in common though: They do not yield a redshift distribution directly but instead some noisy representation of this distribution that needs to be interpreted via a model. This model can either be based on a different calibration approach \citep[like the colour-based techniques discussed above; see e.g.][]{hoyle/etal:2018} or take the free form of a parametric function \citep[spline, Gaussian process, etc.; see e.g.][]{johnson/etal:2017}. Fitting this model to the clustering-$z$ measurements thereby yields a redshift distribution estimate which can be propagated (along with relevant uncertainties) into a cosmological measurement. Here, we follow the methodology laid out in \citet{jlvdb/etal:2020}, who test clustering-$z$ measurements  on mock catalogues that resemble the KiDS+VIKING-450 data set \citep{wright/etal:2019}.

This paper is part of a series of KiDS-1000 papers describing the shear catalogue \citep{giblin/etal:2020}, the methodology behind the cosmological analyses \citep{joachimi/etal:2020}, results from cosmic shear \citep{asgari/etal:2020}, a combined-probes analysis using cosmic shear, galaxy-galaxy lensing, and galaxy clustering from KiDS and BOSS data \citep{heymans/etal:2020}, as well as constraints on cosmological models beyond $\Lambda$CDM \citep{troester/etal:2020}.
Here, we present the redshift distributions used for the cosmological analyses of KiDS-1000. The structure is as follows. In Sect.~\ref{sec:data}, we describe the KiDS-1000 data set, the spectroscopic reference samples, and the mock catalogues that mimic those samples. Section~\ref{sec:SOM} presents results from the SOM method as applied to the KiDS-1000 data and the simulations, and Sect.~\ref{sec:clustering-z} shows how the clustering-$z$ technique is used to further calibrate the redshift distributions based on the SOM method. The results are discussed and the paper is summarised in Sect.~\ref{sec:discussion}, also explaining links to the KiDS-1000 companion papers.

\section{Data}
\label{sec:data}
\subsection{KiDS+VIKING imaging data}

The KiDS-1000 catalogues used here are based on the fourth data release of KiDS \citep[DR4;][]{kuijken/etal:2019}, which includes near-infrared (NIR) photometry based on imaging from the fully overlapping VISTA Kilo degree INfrared Galaxy Survey \citep[VIKING;][]{edge/etal:2013,venemans/etal:2015}. This data set comprises PSF-corrected nine-band $ugriZYJHK_\mathrm{s}$ photometry \citep{kuijken:2008} and BPZ \citep[Bayesian Photometric Redshift;][]{benitez:2000} photo-$z$ estimates for more than 100 million objects over an area of $\sim1000$\,deg$^2$. This constitutes roughly three quarters of the final KiDS+VIKING data set and more than a doubling of the data volume compared to the third data release of KiDS \citep[KiDS-DR3;][]{dejong/etal:2017} that was based on $\sim450$\,deg$^2$ and was used for previous KiDS cosmology analyses \citep[][hereafter H20]{hildebrandt/etal:2020}.

\begin{table*}
  \centering
  \caption{\label{tab:bins}Properties of the five tomographic bins and the full source sample.}
  \begin{tabular}{ccrrrrrrr}
    \hline
    \hline
    bin & selection & \multicolumn{1}{c}{$N$} & \multicolumn{1}{c}{$n_\mathrm{eff}$} & \multicolumn{1}{c}{$\sigma_\epsilon$} & \multicolumn{1}{c}{$N_\mathrm{gold}$} & \multicolumn{1}{c}{$n_\mathrm{eff,gold}$} & \multicolumn{1}{c}{$\sigma_{\epsilon,\mathrm{gold}}$} & \multicolumn{1}{c}{$n_\mathrm{eff,gold}/n_\mathrm{eff}$}\\
    & & & [arcmin$^{-2}$]& & [arcmin$^{-2}$]& \\
    \hline
1   & $0.1<z_\mathrm{B}\le0.3$ &  2\,814\,395 & 0.90 & 0.277 &  1\,792\,136 & 0.62 & 0.270 & 0.69 \\
2   & $0.3<z_\mathrm{B}\le0.5$ &  5\,612\,329 & 1.62 & 0.268 &  3\,681\,319 & 1.18 & 0.258 & 0.73 \\
3   & $0.5<z_\mathrm{B}\le0.7$ &  8\,184\,940 & 2.28 & 0.278 &  6\,148\,102 & 1.85 & 0.273 & 0.81 \\
4   & $0.7<z_\mathrm{B}\le0.9$ &  5\,797\,140 & 1.53 & 0.261 &  4\,544\,395 & 1.26 & 0.254 & 0.82 \\
5   & $0.9<z_\mathrm{B}\le1.2$ &  5\,394\,916 & 1.37 & 0.272 &  5\,096\,059 & 1.31 & 0.270 & 0.95 \\
1-5 & $0.1<z_\mathrm{B}\le1.2$ & 27\,803\,720 & 7.66 & 0.272 & 21\,262\,011 & 6.17 & 0.265 & 0.80 \\
all & ---                     & 31\,446\,584 & 8.43 & 0.273 & N/A          & N/A  & N/A   & N/A  \\
    \hline
  \end{tabular}
\tablefoot{
Effective number densities are calculated with equation~C.12 of \citet{joachimi/etal:2020}, which itself is based on equation~1 of \citet{heymans/etal:2012}. The columns with the gold label correspond to the selection described in Sect.~\ref{sec:SOM}.
}
\end{table*}

Shapes are measured with the \emph{lens}fit software for $\sim31$ million galaxies covering an effective unmasked area of 777.4\,deg$^2$ with a weighted number density of 8.43\,arcmin$^{-2}$ \citep{giblin/etal:2020}. This is the sample used for WL measurements and will be referred to as \emph{sources} in the following. An in-depth description of a very similar sample of roughly half the size called KiDS+VIKING-450 (or KV450) and based on KiDS-DR3 can be found in \citet{wright/etal:2019}. There, the properties of the nine-band photo-$z$ are described in detail and quantified by comparisons to deep spectroscopic redshift catalogues that overlap with KiDS. This information still applies to the KiDS-1000 data used here, as the depth and seeing distributions of KiDS-DR3 and KiDS-DR4 are extremely similar \citep[see ][]{dejong/etal:2017,kuijken/etal:2019}.

The photo-$z$ point estimates $z_\mathrm{B}$, corresponding to the peaks of the posterior redshift distributions of individual galaxies, are used to bin the sources into five tomographic redshift bins. In line with \citetalias{hildebrandt/etal:2020} the first four bins are spaced by $\Delta z_\mathrm{B}=0.2$ in the range $0.1<z_\mathrm{B}\le 0.9$ whereas the fifth bin covers the high photo-$z$ range $0.9<z_\mathrm{B}\le 1.2$. The number densities of the galaxies \citep[according to the definition of][]{heymans/etal:2012} in the five bins are listed in Table~\ref{tab:bins} \citep[for an updated $n_\mathrm{eff}$ estimator that accounts for the impact of the shear responsivity correction, see appendix C of][]{joachimi/etal:2020}.

\subsection{Spectroscopic calibration samples}

The different calibration techniques require spec-$z$ reference catalogues with different properties. For the colour-based calibration, it is required that the reference catalogue spans the same hyper-volume in nine-dimensional magnitude-space, whereas for the clustering-$z$ calibration, a spatially overlapping large-area sample with an extended redshift distribution is needed.

\subsubsection{Deep spectroscopy for colour-based calibration}

The deep spectroscopic sample for the colour-based calibration with the SOM technique (Sect.~\ref{sec:SOM}) did not change between DR3 and DR4.
It consists of a diverse combination of data from the zCOSMOS \citep{lilly/etal:2007,lilly/etal:2009}, VVDS-Deep \citep[VIMOS VLT Deep Survey;][]{lefevre/etal:2005,lefevre/etal:2013,lefevre/etal:2015}, and DEEP2 \citep{newman/etal:2013} projects as well as some additional redshifts from the GAMA \citep[Galaxy And Mass Assembly;][]{driver/etal:2011} deep field G15Deep \citep{kafle/etal:2018} and the CDFS \citep[Chandra Deep Field South; ESO spec-$z$ compilation consisting of spectra from][]{vanzella/etal:2008,popesso/etal:2009,balestra/etal:2010,lefevre/etal:2013}. The main properties of the samples are reported in Table~1 of \citetalias{wright/etal:2020}.

All of these fields have been observed in the nine KiDS+VIKING bands to at least KiDS+VIKING depth, in some cases much deeper. The only exception is the COSMOS field that has no VISTA $Z$-band data. However, it has very deep CFHT (Canada France Hawaii Telescope) $z$-band data \citep{hildebrandt/etal:2009a}, which, due to the similarity of the MegaCam@CFHT $z$-band and the VIRCAM@VISTA $Z$-band, can be used as a substitute. In cases where the imaging data in the deep redshift calibration fields is deeper than in KiDS+VIKING, we added Gaussian noise to arrive at a data set that is representative for KiDS+VIKING. In principle, one could also make use of deeper data in the calibration fields and improve the precision of the calibration for instance as described by \citet{buchs/etal:2019}, but we leave such an enhancement of the KiDS+VIKING redshift calibration to future work.

\subsubsection{Wide-area spectroscopy for clustering redshifts}
\label{sec:wide_ref_K1000}
In \citetalias{hildebrandt/etal:2020}, clustering-$z$ (CZ)\footnote{We note that clustering-$z$ were abbreviated as CC (cross-correlations) in previous KiDS papers. Here, we opt to switch to the new acronym CZ to more specifically refer to clustering-$z$.} were estimated with the help of spec-$z$ data from the wide-area surveys GAMA-DR3 \citep{baldry/etal:2018}, SDSS-DR12 \citep{eisenstein/etal:2011,alam/etal:2015}, 2dFLenS \citep{blake/etal:2016}, and WiggleZ \citep{drinkwater/etal:2010} and complemented with information about the high-redshift part of the $n(z)$ from zCOSMOS, VVDS-Deep, and DEEP2. The same samples are employed here but with some significant changes, the most important one being approximately a doubling in the size of the overlap area between KiDS+VIKING and SDSS in the Northern Hemisphere as well as between KiDS+VIKING and 2dFLenS in the Southern Hemisphere. This alone significantly increases the signal-to-noise ratio (S/N) of the CZ measurements as described in Sect.~\ref{sec:clustering-z}. Additionally, we have relaxed some of the very conservative masking in previous KiDS CZ analyses.

From the SDSS spec-$z$ compilation we only use sources observed as part of BOSS \citep[Baryon Oscillation Spectroscopic Survey;][]{dawson/etal:2013} unlike in previous KiDS work where also the SDSS Main Galaxy Sample \citep[MGS;][]{strauss/etal:2002} and the SDSS Quasar Sample \citep{schneider/etal:2010} were used. The reason behind this decision is the desire to minimise systematic errors through the correction for evolving galaxy bias, which becomes more complicated when different samples are combined. At low redshift, we have very high S/N from GAMA already and do not need the limited additional information from the SDSS-MGS. While a higher S/N at high redshift would be desirable, the sparsity of the SDSS-QSO sample does not add any significant information and the results are almost indistinguishable whether it is included or not.

The spec-$z$ samples used for the CZ measurements are summarised in Table~\ref{tab:speccluster}. We note that the areas in the COSMOS and VVDS-Deep fields used for CZ are slightly smaller than those used for the colour-based calibration as the former has stricter requirements on the spatial homogeneity of the data.

\begin{table}
\centering
\caption{Spectroscopic redshift samples used for the clustering-$z$ calibration.}\label{tab:speccluster}
\begin{tabular}{lrr}
\hline
\hline
Survey & No. of & Area\tablefootmark{a} \\ 
     & spec-$z$ & $[\rm deg^2]$ \\
\hline
zCOSMOS  &   $8\,422$ &   $0.5$ \\
DEEP2    &   $8\,698$ &   $0.8$ \\
VVDS     &   $4\,194$ &   $0.5$ \\
GAMA     & $114\,912$ & $137.4$ \\
BOSS     &  $47\,332$ & $262.5$ \\
2dFLenS  &  $17\,231$ & $266.1$ \\
WiggleZ  &  $42\,328$ & $130.1$ \\  
\hline
total    & $321\,318$ & $784.8$\tablefootmark{b} \\
\hline
\end{tabular}
\tablefoot{
\tablefoottext{a}{The area quoted for the wide fields is a rough estimate calculated from the number of pointings that go into each cross-correlation measurement and the average unmasked area per pointing.}
\tablefoottext{b}{We note that there is significant overlap between GAMA, BOSS, and WiggleZ. Hence, the total area quoted here is not to be understood as an independent area.}
}
\end{table}

\subsection{MICE mock catalogues}
\label{sec:MICE}

The KiDS+VIKING redshift calibration is validated on simulated mock catalogues based on the MICE simulation \citep{fosalba/etal:2015a,fosalba/etal:2015b,crocce/etal:2015,carretero/etal:2015,hoffmann/etal:2015}. The creation and properties of these mock catalogues is covered in detail in \citet{jlvdb/etal:2020}. The KiDS+VIKING nine-band photometry and the BPZ photo-$z$ are simulated within these mocks, whereas the shape measurement weights are sampled from the real data by assigning each mock galaxy the weight of its nearest neighbour in the KiDS-1000 data in $r$-band magnitude. This results in a mock source catalogue that closely resembles the data. The most important difference is that MICE only provides mock galaxies out to $z\sim1.4$. Hence, we cannot test for possible high-$z$ tails with the help of this mock, but we note that the core of the redshift distribution of each tomographic bin is well covered by these mocks.

In a similar way, the spec-$z$ calibration samples are simulated by applying the original (or in some cases slightly modified) selection criteria to the mock photometry and implementing realistic magnitude- and redshift-dependent spectroscopic success rates. For details, we refer the reader to \citet{jlvdb/etal:2020}. We also create an idealised reference sample by taking every 10th KiDS mock source. This somewhat unrealistic case can be used to test the CZ methodology and explore the unavoidable systematic error floor inherent to our CZ implementation, agnostic to the complexities of reference sample construction. 

The mock catalogues for the deep spectroscopic fields are identical to the ones used in \citetalias{wright/etal:2020}. Hence, the mock results for the SOM calibration from \citetalias{wright/etal:2020} also apply to the data set presented here. These results will be discussed in Sect.~\ref{sec:SOM}.

The mock catalogues for the CZ measurement are simply expanded in area compared to the ones in \citet{jlvdb/etal:2020} to account for the larger area of the KiDS-1000 source sample compared to KV450. In fact, for the analysis presented here we create mock catalogues for all samples (WL sources, deep spec-$z$ surveys, wide spec-$z$ surveys) over an area of 744.4\,deg$^2$ split into 1024 pointings of 0.727\,deg$^2$ each. In particular for the deep fields, having such a large number of realisations makes it possible to estimate covariance matrices from the simulations that can be used to combine the results from the different surveys on the real data. One notable difference to the mock catalogues presented in \citet{jlvdb/etal:2020} is the fact that we use a pure BOSS sample instead of a combined SDSS sample also including the Main Galaxy and QSO samples, mirroring the approach taken on the KiDS-1000 data (see Sect.~\ref{sec:wide_ref_K1000}).

\section{Colour-based redshift calibration with a self-organising map}
\label{sec:SOM}

Photometric redshifts rely on the fact that galaxy colours strongly correlate with redshift. The same information is exploited in the calibration of redshift distributions for WL applications with the help of a deep spec-$z$ reference sample. In essence, this is quite similar to the well-known category of ML photo-$z$, with the important difference that we want to apply this to a target ensemble of galaxies with unknown redshifts rather than to individual galaxies. Additionally, the goals of colour-based $n(z)$ calibration are somewhat different from the goals of most ML photo-$z$ codes, with the former being optimised towards low bias in the mean redshift and the latter often towards low scatter and low outlier rates.

\begin{figure*}
 \centering
 \includegraphics[width=0.95\hsize]{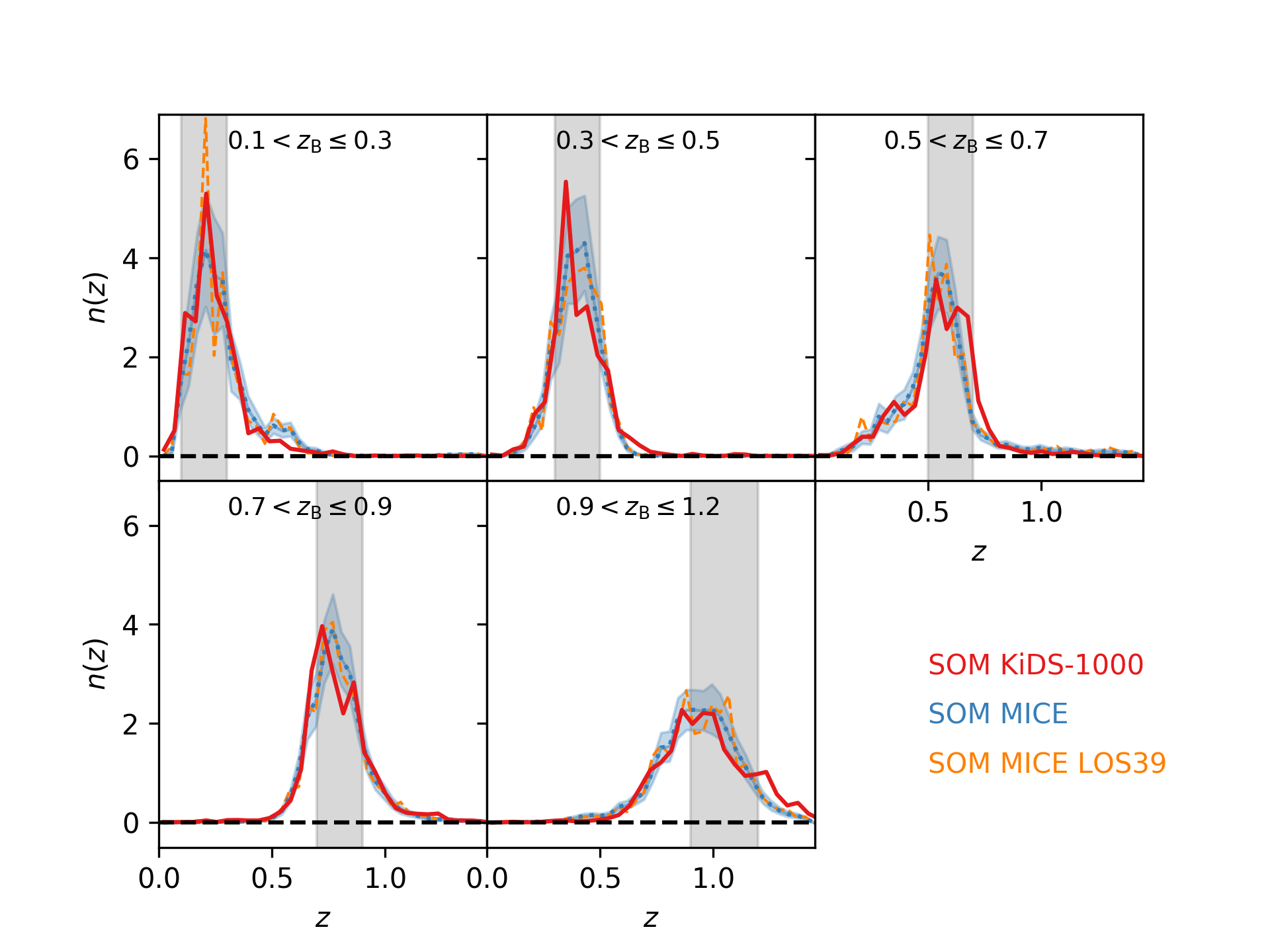}
 \caption{Redshift distributions for the five tomographic redshift bins used in the KiDS-1000 cosmological analyses estimated with the SOM method of \citet{wright/etal:2019}. The grey vertical bands indicate the photo-$z$ cuts defining the bins. Solid red lines show the estimate from the KiDS-1000 data whereas the dotted blue lines and their confidence intervals represent the average and standard deviation of all lines-of-sight of the MICE mocks. The dashed orange lines show one representative (in terms of its mean redshifts) line-of-sight (number 39 in our list) that is used in Sect.~\ref{sec:CZ_res_MICE}. \label{fig:SOMnz}}
\end{figure*}

\subsection{Method}

The inherent differences between a spec-$z$ calibration sample and a typical WL source sample can - under certain circumstances - be overcome by re-weighting. This re-weighting of the calibration sample is supposed to make the distributions of relevant quantities as similar as possible between the two samples. Once this is achieved, it is assumed that the weighted distribution of spec-$z$ in the calibration sample should be a good estimate of the unknown distribution of redshifts of the target source sample. It is clear that this works better the more spec-$z$ are available, the more complementary information (e.g. number of photometric bands) is used to establish the weighting, and the closer the selection of the spec-$z$ sample resembles the source sample to start with \citep{gruen/etal:2017}.

\citet{lima/etal:2008} suggest an approach that estimates the density of both samples in high-dimensional magnitude-space via a $k$-nearest-neighbour ($k$NN) method. The ratio of the densities in each point in magnitude-space is then used as a weight for the spec-$z$ in that place. Essentially, spec-$z$ that are underrepresented compared to the unknown target sample are up-weighted and spec-$z$ that are over-represented are down-weighted. It can be shown on simulations \citep[][]{lima/etal:2008,wright/etal:2020} that this approach yields good results if the magnitude-space is sufficiently high-dimensional, the photometry has high S/N, and the magnitude-space of the target sample is fully covered with spec-$z$ calibrators.

 Whether the first two requirements are sufficiently met can realistically only be investigated with simulations. The third requirement, however, can partly be assessed with the data themselves by checking the overlap of the target and calibration samples. Being a high-dimensional problem, such checks need to make use of some dimensionality-reduction technique. \citet{masters/etal:2015} argued that SOMs are well suited for this purpose.

\citetalias{wright/etal:2020} show that the SOM method can be used to actually carry out the estimation of the redshift distribution, $n(z)$, without further need for the $k$NN method. Instead of estimating the densities of target and calibration sample at the location of each calibration source, the densities are estimated in each cell of the SOM. Moreover, the SOM gives the user a simple tool to cull from the target sample sources that are not represented by the spec-$z$ calibration sample, that is sources that lie in cells that are not filled with at least one reference object. Each tomographic redshift bin is calibrated individually with a calibration sample that is limited to the same photo-$z$ ($z_\mathrm{B}$) range. The following additional criterion is established to select good SOM cells:
\begin{equation}
  | \langle z^\mathrm{s}_\mathrm{spec}\rangle_i - \langle z ^\mathrm{p}_\mathrm{B}\rangle _i | < \mathrm{max}\left[ 5 \times \mathrm{nMAD}\left( \langle z ^\mathrm{s}_\mathrm{spec}\rangle  - \langle z ^\mathrm{s}_\mathrm{B}\rangle \right), 0.4 \right]\,, \label{eq:QC1}
\end{equation}
where the superscripts $\rm s$ and $\rm p$ refer to the spectroscopic calibration and the photometric target samples, respectively, the angular brackets indicate an unweighted average, the index $i$ refers to a single SOM cell, and the normalised median absolute deviation\footnote{Normalised in such a way that it equals the standard deviation for a Gaussian distribution.} on the right-hand side is taken over the full SOM. This criterion rejects cells that show suspiciously large deviations between the mean spectroscopic redshift of all calibration objects in a cell and the mean photo-$z$ of all target objects in that cell. See \citetalias{wright/etal:2020} and \citet{wright/etal:2020b} for more details.

In this way, \citetalias{wright/etal:2020} define a KiDS `gold' sample with smaller number density and more robust $n(z)$ estimates. The KiDS-1000 WL analyses all make use of this gold selection to benefit from the robustness of the $n(z)$ estimates.

\citet{wright/etal:2020b} analyse the gold sample for KV450 with the corresponding SOM-based $n(z)$, finding very good agreement in their cosmological parameter estimates with previously published results based on the full samples and an $n(z)$ estimated with the $k$NN method \citepalias{hildebrandt/etal:2020}.
The SOM analysis presented here follows the methods presented in \citetalias{wright/etal:2020}. Given that the spec-$z$ calibration sample is identical in both studies, the only difference is the larger (by a factor of $\sim2$) target catalogue with slightly updated absolute photometric calibration and updated \emph{lens}fit weights \citep{giblin/etal:2020}. As the SOM analysis was not limited by the (already large) size of the target sample in \citetalias{wright/etal:2020}, this should only result in very minor changes to the $n(z)$.

\subsection{Results from MICE mocks}
\label{sec:SOM_res_MICE}
\citetalias{wright/etal:2020} use the MICE mock catalogues described in Sect.~\ref{sec:MICE} to estimate residual biases in their SOM-estimated redshift distributions. After optimising the SOM setup with a series of tests they also introduce additional clustering of the SOM cells.\footnote{This is not to be confused with the physical clustering of galaxies and just describes the merging of SOM cells with similar properties.} By combining multiple cells into a cluster, an optimal compromise between fidelity and shot-noise is found. The redshift distributions estimated with the SOM technique on the mock catalogues are displayed in Fig.~\ref{fig:SOMnz}.

\citetalias{wright/etal:2020} report values for the bias of the mean redshift in the five tomographic bins used for the gold cosmic shear analysis of \citet{wright/etal:2020b} in the form
\begin{equation}
 \label{eq:dzSOM}
 \Delta\langle z\rangle^{\rm SOM}_j = \langle z\rangle^{\rm SOM}_j - \langle z\rangle^{\rm true}_j \,,
\end{equation}
where the averages are taken per tomographic bin $j \in \{1,2,3,4,5\}$.
Since the mocks did not change in the meantime and the KiDS-1000 data closely resemble the KV450 data, the same biases apply to the KiDS-1000 calibration presented here. We note that the improvement to the \emph{lens}fit weight recalibration methodology between KV450 and KiDS-1000, as discussed in section~2.2 of \citet{giblin/etal:2020}, is not propagated into the mock catalogues as it does not significantly change the mean properties of the KiDS-1000 tomographic bins compared to the KV450 bins. Values for the mean biases and their uncertainties as estimated from 100 simulated lines-of-sight are reported in the second column of Table~\ref{tab:results}. Those can be compared to the biases estimated for the mean redshifts of the full samples of \citetalias{hildebrandt/etal:2020} with the $k$NN method \citepalias[last line of Table~3 of][]{wright/etal:2020}, which are significantly larger and range from $0.047$ in the first bin to $-0.013$ in the fifth bin. 

\begin{table*}
 \centering
 \caption{\label{tab:results}Redshift calibration for the five tomographic bins used in the KiDS-1000 cosmology analyses. }
 \begin{tabular}{crrrr}
   \hline
   \hline
   bin & $\Delta \langle z\rangle^{\rm SOM}$ & $\delta z^{\rm CZ} \pm \text{stat.} \pm \text{syst.}$ & $\delta z^{\rm CZ} \pm \text{stat.} \pm \text{syst.} $ & $\delta z^{\rm CZ} \pm \text{comb.}$\\
       & MICE & MICE & KiDS & KiDS \\
   \hline
1 & $ 0.000\pm0.011$ & $ 0.001\pm0.002\pm0.004$ & $-0.001\pm0.004\pm0.004$ & $-0.001\pm0.012$ \\
2 & $ 0.002\pm0.011$ & $-0.002\pm0.002\pm0.004$ & $ 0.004\pm0.003\pm0.005$ & $ 0.004\pm0.013$ \\
3 & $ 0.013\pm0.012$ & $ 0.004\pm0.003\pm0.010$ & $ 0.011\pm0.004\pm0.016$ & $ 0.011\pm0.020$ \\
4 & $ 0.011\pm0.009$ & $ 0.015\pm0.001\pm0.024$ & $-0.008\pm0.006\pm0.007$ & $-0.008\pm0.013$ \\
5 & $-0.006\pm0.010$ & $ 0.003\pm0.002\pm0.004$ & $ 0.003\pm0.007\pm0.003$ & $ 0.003\pm0.013$ \\
 \end{tabular}
\tablefoot{Bias in the mean redshift (column 2) as estimated with the SOM method from the MICE mocks \citepalias{wright/etal:2020}. The uncertainties have been multiplied by a factor of two to account for residual differences between mocks and data. Columns 3~\&~4 report the best-fit values for the $\delta z^{\rm CZ}$ parameters (defined in Eq.~\ref{eq:dzCZ}) on the MICE mocks and the KiDS-1000 data, respectively. The values from column 4 are based on fits to the SOM redshift distributions, which carry their own uncertainty (column 2). In column 5 we report the same shifts as in column 4 but combine all sources of uncertainty.}
\end{table*}

\begin{figure}
 \centering
 \includegraphics[width=0.71\hsize]{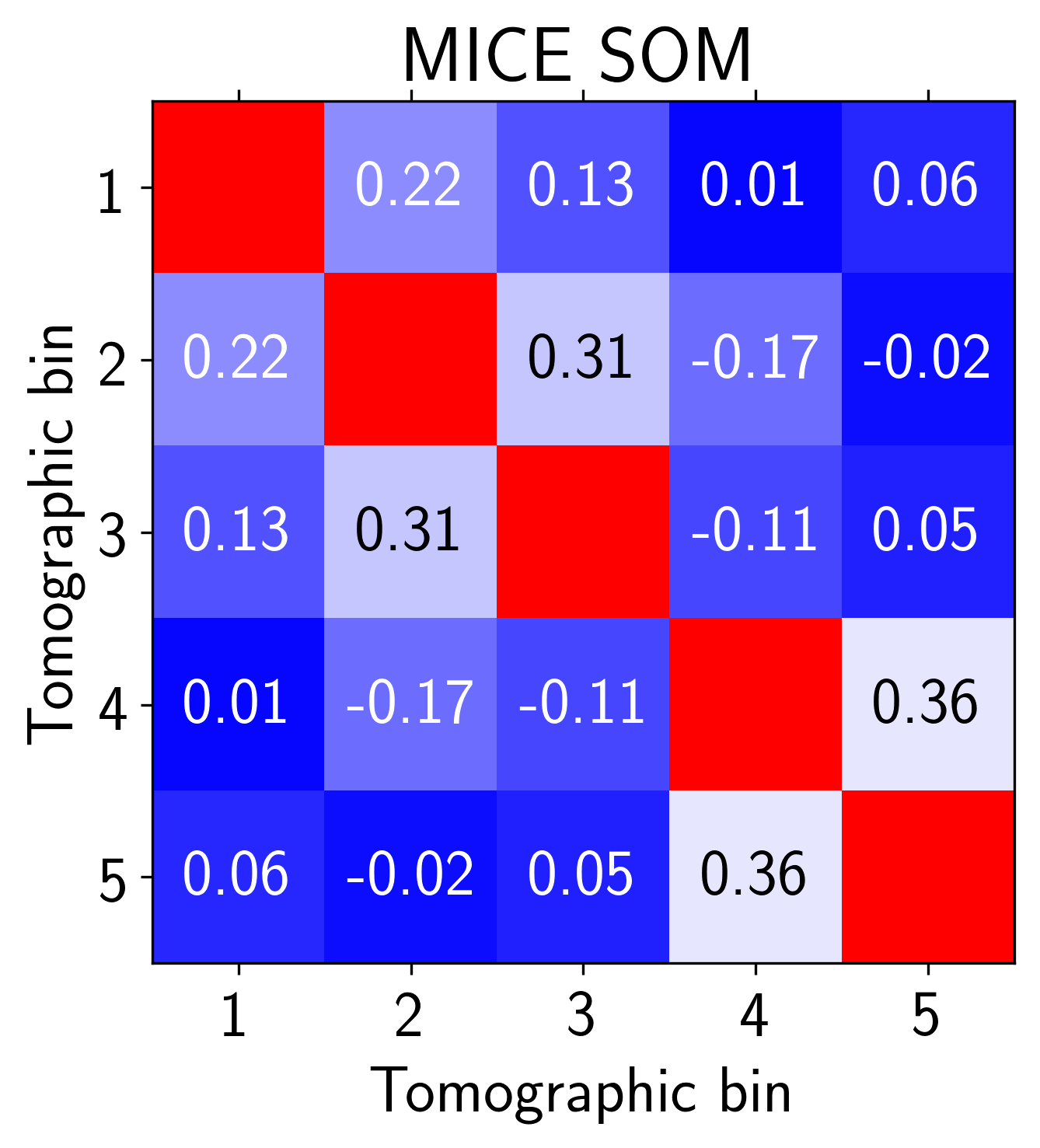}
 \caption{\label{fig:MICE_corr}Correlation matrix of the uncertainties of the $\Delta \langle z\rangle^{\rm SOM}_i$ from the SOM analysis of the MICE mocks reported in column 2 of Table~\ref{tab:results}.}
\end{figure}

As the uncertainties quoted in Table~\ref{tab:results} are estimated from 100 realisations along different lines-of-sight for the mock spec-$z$ calibration sample, these uncertainties include contributions from photometric noise, shot-noise due to the limited sample size, spectroscopic selection effects and incompleteness, and sample variance due to large-scale structure. The latter effect leads to a correlation of the uncertainties, which is also estimated from these 100 realisations. We report the correlations in Fig.~\ref{fig:MICE_corr}. Neighbouring tomographic bins are correlated by up to 36\%, while more widely separated bins are only weakly correlated or also weakly anti-correlated.

The uncertainties and their correlations are taken into account in the cosmological analyses with the KiDS-1000 data \citep{asgari/etal:2020,heymans/etal:2020,troester/etal:2020}. In order to account for inherent imperfections in the simulation we conservatively enlarge all these uncertainties by a factor of two in the fiducial analyses.

\subsection{Results from KiDS-1000 data}

We update the SOM analysis of \citetalias{wright/etal:2020} by populating the SOM with the new KiDS-1000 catalogues instead of the KV450 catalogues that were used in that paper. This leads to slightly different redshift distributions, which are also displayed in Fig.~\ref{fig:SOMnz}, and different effective number densities ($n_\mathrm{eff}$) as well as ellipticity dispersions $\sigma_\epsilon$ reported in Table~\ref{tab:bins}.
By applying the gold selection, roughly 20\% of the effective source density is removed, which slightly increases the statistical noise (shape noise). Partly counteracting the decrease in $n_\mathrm{eff}$ , however, is a small reduction in $\sigma_\epsilon$, which sets the noise level per source of WL measurements.

Comparing the number densities of the gold selection for KV450 and KiDS-1000 (see Table~\ref{tab:bins} here and Table~2 of \citetalias{wright/etal:2020}) reveals some notable differences. In particular, the first and second tomographic bins show significantly lower representation fractions on the KiDS-1000 data. We attribute this to subtle differences in the absolute photometric calibration between KV450 and KiDS-DR4 \citep[see][]{wright/etal:2019,kuijken/etal:2019} combined with our assumed number of hierarchical clusters in the SOM. \citetalias{wright/etal:2020} demonstrate that, while the choice of cluster number (see their figure B1, panel [b]) can introduce swings of $>20\%$ in representation fraction, the reconstructed redshift distributions remain entirely unbiased (panel [c]). As a result of this conclusion, we chose not to re-optimise the number of hierarchical clusters used for DR4 even after our slight change in calibration and retraining of the fiducial SOM.

\section{Calibration with clustering redshifts}
\label{sec:clustering-z}
In the following, we describe the complementary clustering redshift technique that yields an independent estimate of the mean redshifts of the galaxies in the tomographic bins.

\subsection{Method}
\label{sec:CZ_method}

\subsubsection{Measurement}
The clustering redshift methodology used for KiDS-1000 closely follows the approach described in \citet{jlvdb/etal:2020}. This approach implements the technique suggested by \citet{schmidt/etal:2013} using small-scale clustering in a single broad radial bin with an additional radial weighting. By pre-selecting galaxy samples that are already relatively narrowly located in redshift (i.e. the tomographic redshift bins), any effects of evolving galaxy bias are minimised to start with. The evolution of the galaxy bias of the reference sample is mitigated by estimating the angular auto-correlation function of this sample in the same radial- and redshift bins as the cross-correlation measurement. Any residual effect of the bias evolution of the source sample itself can in principle be mitigated via an internal consistency check \citep[also called self-consistency bias mitigation or SBM; see Sect.~3.3 of][]{jlvdb/etal:2020}. This check is based on comparing the results from a broad target sample (e.g. all tomographic bins combined) with the weighted sum of the narrow samples. As each of the narrow samples gets normalised individually, the weighted sum is not exactly equal to the measurement on the broad sample. Differences can be interpreted as being due to evolving galaxy bias and approximated by a parametric model \citep{davis/etal:2018}.

However, the galaxy bias of typical WL source samples evolves only very slightly over the redshift baseline of the core of a single tomographic bin. This limits the importance and usefulness of this approach. Only at high S/N of the CZ measurements can this additional complexity in the model be constrained by the data. We distinguish the following estimates of the redshift distribution:
\begin{align}
 w_{\rm CZ}(z) & \hspace{0.5cm}\text{raw CZ measurements}\label{eq:w}\\
 \tilde n_{\rm CZ}(z) & \hspace{0.5cm}\text{CZ after correction for reference bias}\label{eq:ntilde}\\
 n_{\rm CZ}(z) & \hspace{0.5cm}\text{fully corrected CZ (reference bias and SBM)}\,,\label{eq:n}
\end{align}
\citep[see][for the performance of the different options on mock catalogues]{jlvdb/etal:2020}. We note that the bias of outlier populations with a redshift very different from the core of the $n(z)$, and potentially also a linear bias value that is very different, cannot be reliably corrected with this method. As such, this method is only useful for narrow, unimodal redshift distributions.

In comparison to previous KiDS analyses, we have implemented a number of changes to the CZ methodology. Unlike \citet{hildebrandt/etal:2017,hildebrandt/etal:2020} we use an updated version of the angular cross-correlation code \texttt{the-wizz} \citep{morrison/etal:2017} called \texttt{yet-another-wizz} or \texttt{yaw}. We refer the reader to \citet{jlvdb/etal:2020} for a detailed description of the features of \texttt{yaw}. The main advantage of this new version is that it avoids the inherent sky pixelisation of \texttt{the-wizz}, which is inherited from the library \texttt{STOMP} \citep{scranton/etal:2002}. This improvement yields more realistic uncertainties, especially for small angular scales that are often probed at high redshift for a given comoving scale.

We further experiment with different radial scales. \citet{jlvdb/etal:2020} used comoving scales of $100\,\mathrm{kpc}<r<1\,\mathrm{Mpc}$ for their measurements throughout. Here we also explore the performance of the CZ method with additional scales of  $30\,\mathrm{kpc}<r<300\,\mathrm{kpc}$, $50\,\mathrm{kpc}<r<500\,\mathrm{kpc}$, and $500\,\mathrm{kpc}<r<1.5\,\mathrm{Mpc}$.\footnote{We note that a cosmological model needs to be assumed to convert angular scales into comoving distances. Here, we assume a Planck-2015  cosmology \citep{planck2015}, but this choice has negligible influence on our results as long as the same scales are used consistently for all correlation function measurements of a given tomographic bin.} Especially, the smaller scales yield very high S/N, at the price of potentially more complex bias evolution. With the mock catalogues, the impact of this can be tested. We note that the limited CZ analysis of \citet{hildebrandt/etal:2017} also measured over scales of $30\,\mathrm{kpc}<r<300\,\mathrm{kpc}$ and reached a usable S/N from less than $2\,\mathrm{deg}^2$ of area covered by deep pencil-beam surveys.

The redshift binning is less critical as our approach should be able to correct for all biases regardless of this binning. Here we choose 45 redshift bins of constant radial comoving length in the redshift range $0<z<3$. We use the same binning for the data and the mocks but can essentially only use the lower half of the redshift range for the mocks as the MICE galaxy population only extends to $z=1.4$.

\subsubsection{Covariance}
The other update compared to \citet{jlvdb/etal:2020} and the CZ analysis in \citetalias{hildebrandt/etal:2020} concerns the covariance matrix of the CZ measurements. Due to the limited size of the reference samples all previous CZ analyses with KiDS estimated the covariance from a bootstrap or jackknife re-sampling over all ($\sim1$\,deg$^2$) pointings that went into the measurement. We will call this approach of estimating the covariance via bootstrap (A) in the following and use it by default.

This implementation of re-sampling neglects any differences in spectroscopic coverage between pointings. In effect, the subsamples, which the bootstrap samples are constructed from, can have different statistical weights, especially at high redshift. In general, this leads to an underestimation of the uncertainty of the CZ measurements. As shown by the mock analysis of \citet{jlvdb/etal:2020}, which also uses approach (A) and in principle also suffers from the same deficiency, this can still yield sufficiently accurate results. In the following we try to estimate the additional uncertainty due to this effect and propagate it into our results.

Instead of treating all measurements from the different reference surveys equally, we can also split the analysis and first analyse the different surveys independently. While this was not really possible with previous KiDS data releases, even after splitting the KiDS-1000 data volume still leaves $>100$ pointings for each of the wide-area reference surveys $i$ to empirically estimate the CZ data $n_i(z)$ and a corresponding covariance matrix $\tens{C}_i$ via bootstrap re-sampling. The measurements of all wide-area surveys are then combined with precision weighting (or inverse covariance weighting) assuming uncorrelated Gaussian uncertainties
\begin{equation}
\label{eq:pw_data}
n(z) = \tens{C} \sum_i\tens{C}_i^{-1}\vec{n}_i\,,
\end{equation}
where $\tens{C}$ is the combined, precision-weighted covariance estimated as
\begin{equation}
\label{eq:pw_cov}
\tens{C} = \left(\sum_i\tens{C}_i^{-1}\right)^{-1}
\end{equation}
and $\vec{n}_i$ is the redshift distribution vector of the $i$th bin.
This approach will be called (B) in the following and it can be applied to any of the three estimates for CZ described in Eqs.~\ref{eq:w}-\ref{eq:n}. In this way, only subsamples with comparable statistical properties enter each of the bootstrap estimates, making those more reliable. However, as mentioned above, this method also assumes that there is no correlation between the measurements from different reference samples, which is not true due to the overlap of some of these samples (GAMA, BOSS, WiggleZ). Again, we test the impact of violating this assumption on the mocks, which replicate the overlap of the reference surveys in the KiDS-1000 data. Due to the small data volume this approach (B) is not feasible on the deep pencil-beam fields.

We complement these two empirical estimates of the CZ covariance with a simulation-based approach that we will call (C) in the following. Instead of applying any bootstrap re-sampling to the data, we leverage the MICE mock catalogues described in Sect.~\ref{sec:MICE} to estimate a covariance matrix. For each of the seven reference samples quoted in Table~\ref{tab:speccluster} - regardless of whether it is a wide-area or deep pencil-beam sample - we measure the CZ on 1024 pointings of the MICE mocks. This is sufficient to estimate a low-noise covariance matrix via bootstrap re-sampling for each individual reference sample, which can then be scaled to the actual area quoted in Table~\ref{tab:speccluster}. Measurements and corresponding covariance matrices from the different reference samples are then combined again with precision weighting.

\begin{figure}
\centering
\includegraphics[width=\hsize,clip=true,trim=1.5cm 0cm 1cm 1cm]{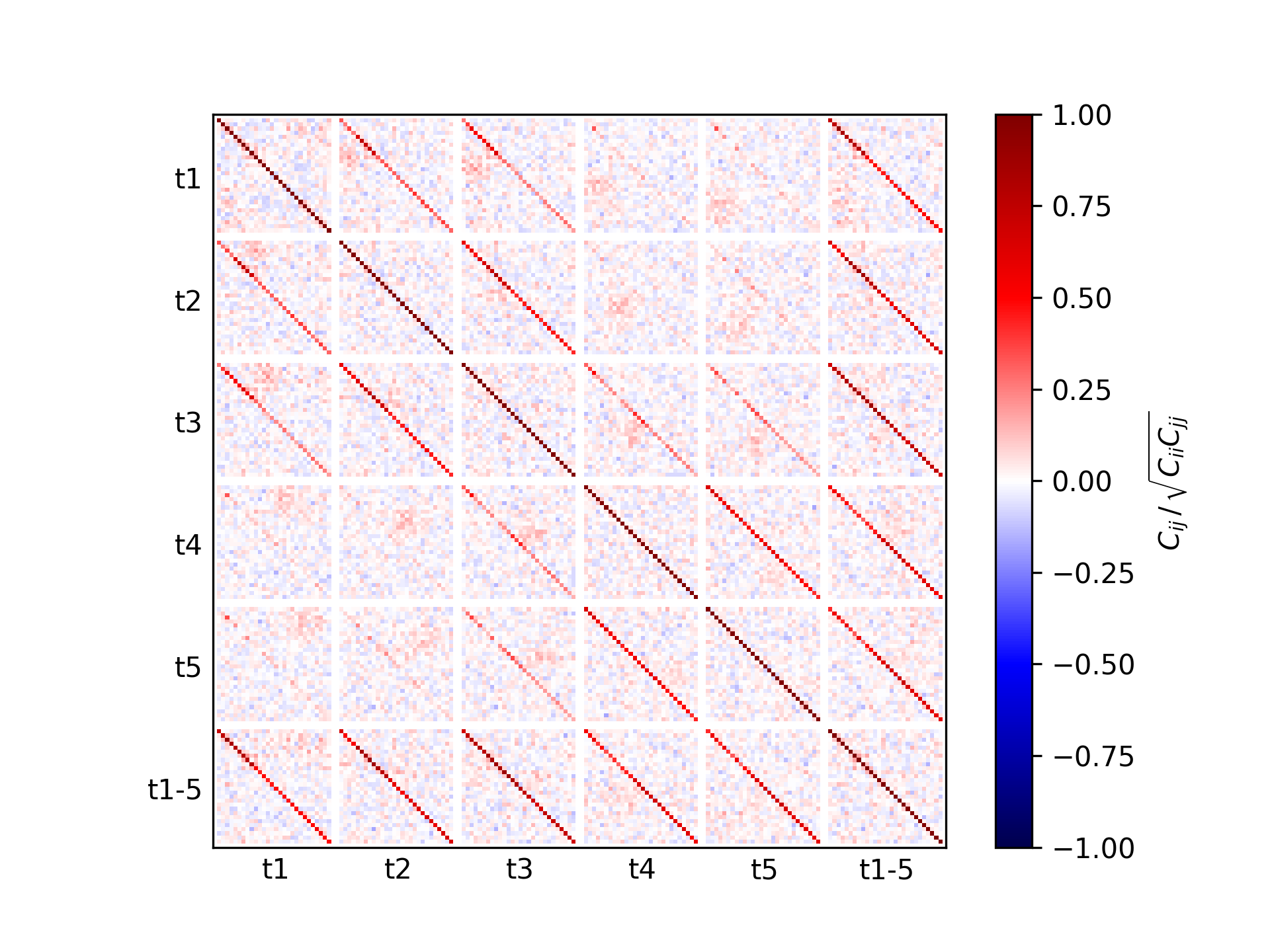}
\caption{\label{fig:corr} Correlation matrix of CZ measurements from the MICE mocks using an idealised reference sample with high number density. There are six blocks in a line, each $30$ pixels wide corresponding to 30 redshift bins in the range $0<z<1.4$ (with the first and last redshift bin containing no galaxies due to the redshift limits of MICE and shown white here). The first five blocks correspond to the five tomographic bins and the sixth block to the combined sample. The latter one is obviously correlated with all other samples as it shares target galaxies with the other bins.}
\end{figure}

All three covariances show consistently that the CZ measurements are not strongly correlated between different redshifts, as expected from uncorrelated large-scale structure along the line-of-sight. See an example correlation matrix of the idealised reference sample cross-correlated with the target tomographic bins in Fig.~\ref{fig:corr}. We always suppress noise in the covariance by setting those covariance elements to zero that correspond to different redshift bins. However, at a given redshift there is some correlation between the measurements for different tomographic bins (most pronounced for neighbouring bins) as these measurements are based on the same reference objects.

The different approaches to estimate the covariance are affected by different levels of noise, with approach (A) on the deep pencil-beam surveys being the noisiest and approach (C) generally being the least noisy. Depending on the noise level we decide whether to use or ignore the off-diagonal elements that correlate measurements in different tomographic bins at the same redshift. For example, with our fiducial approach (A) on the deep fields the estimates of these covariance elements are too noisy and need to be ignored to allow for an inversion of the matrix.

\begin{figure*}
\centering
\includegraphics[width=0.95\hsize]{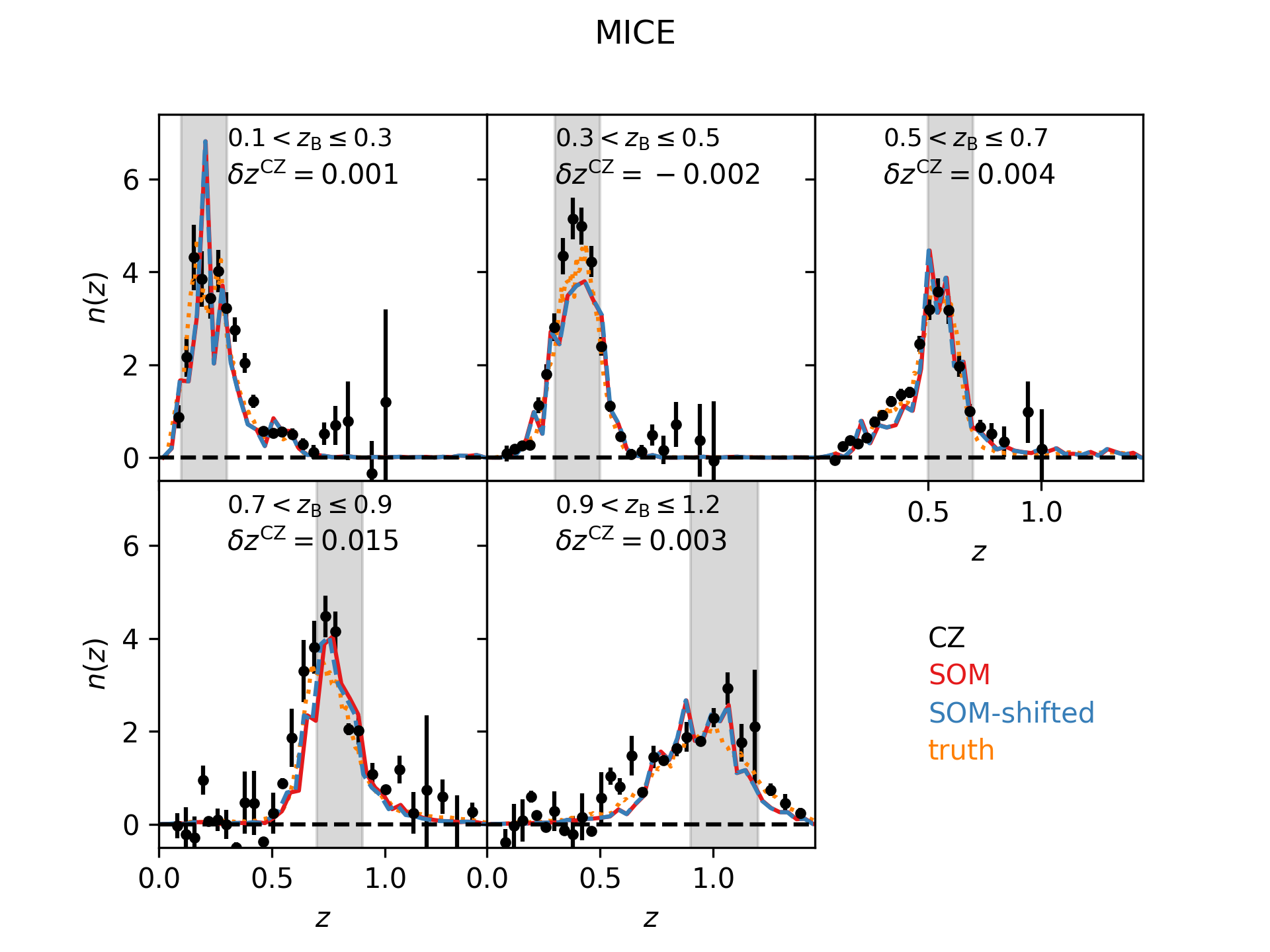}
\caption{\label{fig:res_MICE}Clustering-$z$ measurements on the MICE mocks with the fiducial setup, i.e. using the wide fields and scales of $100\,{\rm kpc}<r<1\,{\rm Mpc}$ for the first three tomographic bins (top row) and the deep fields and scales of $30\,{\rm kpc}<r<300\,{\rm kpc}$ for the upper two tomographic bins (bottom row). The original SOM redshift distributions from a representative line-of-sight are shown in solid red and the best-fit model is shown in dashed blue. The true redshift distributions are shown in dotted orange for comparison.}
\end{figure*}

\subsubsection{Model fitting}
Clustering-$z$ measurements are noisy representations of an underlying redshift probability distribution. Noise fluctuations can lead to negative clustering amplitudes that cannot be readily converted into a probability density. Hence, one needs a model to interpret these noisy data points. In general, we minimise
\begin{equation}
\label{eq:shift_fit}
\chi^2 = \left[ n_{\rm CZ}(z) - m_\theta(z) \right]^T \tens{C}^{-1} \left[ n_{\rm CZ}(z) - m_\theta(z) \right] \, ,
\end{equation}
where $m_\theta(z)$ is some model of the clustering-redshift distribution $n_{\rm CZ}(z)$ with parameters $\theta$.

In \citetalias{hildebrandt/etal:2020} and \citet{jlvdb/etal:2020} we used redshift distributions from the $k$NN re-weighting technique \citep[][dubbed DIR in previous KiDS papers]{lima/etal:2008} as a model, which was shifted by an offset $\delta z^{\rm CZ}$ to yield a best-fit to the CZ data. Here, we switch to the SOM-estimated redshift distributions from Fig.~\ref{fig:SOMnz} as a model, that is
\begin{equation}
\label{eq:dzCZ}
m_\theta(z) = A\, n_{\rm SOM}(z+\delta z^{\rm CZ})\,,
\end{equation}
where $\theta=(A,\delta z^{\rm CZ})$ are the fit parameters to be minimised. We typically only report $\delta z^{\rm CZ}$ as the value of the amplitude $A$ is unimportant after normalisation of the best-fit model. We note that we distinguish between discrete differences in mean redshifts as $\Delta\langle z\rangle^{\rm SOM}$ in Eq.~\ref{eq:dzSOM} and continuous fitting parameters such as $\delta z^{\rm CZ}$ in Eq.~\ref{eq:dzCZ} by the use of capital $\Delta$ and small $\delta$, respectively.

We expect the SOM redshift distributions to be less biased than the DIR-estimated ones \citep{wright/etal:2019} so that the results should come closer to the idealised case discussed in \citet{jlvdb/etal:2020}, where the true redshift distributions were used on the MICE simulations to discover any residual biases in the CZ method. We also report such results from some tests with the true redshift distribution for MICE below.

The motivation for using the SOM $n(z)$ and fitting a shift is the fact that cosmic shear measurements are mostly sensitive to the mean redshift of the source sample. A bias in the mean redshift due to a coherent offset of the core of the redshift distribution is readily captured in the best-fit value of this shift parameter. However, it should be noted that a bias in the mean due to outliers cannot be captured by this simple model.

Another general problem with this approach is that the shape of the DIR or SOM $n(z)$ is not perfectly accurate, meaning their higher-order moments differ from the true redshift distribution. While the DIR $n(z)$ are typically too broad, the opposite is true for the SOM $n(z)$ that are typically slightly too narrow. These properties are revealed by the mock analysis of \citet{wright/etal:2019}. If the S/N of the CZ measurements changes significantly with redshift, such a bias in the shape of the model can lead to a bias in the mean redshift as estimated from the best-fit shift parameter $\delta z^{\rm CZ}$. This can be easily understood by imagining some CZ measurement for a tomographic bin, whose S/N is high on the low-$z$ side of the peak and low on the high-$z$ side. The fit of any model will be driven by the high S/N data points at low-$z$ and influenced little by the low S/N data points at redshifts higher than the peak. If the model is too broad this will bias the inferred mean redshift high, if the model is too narrow the inferred mean redshift will be biased low. See Appendix~\ref{sec:toy_model} for a toy model illustrating this effect.

In order to avoid such problems, one could add more parameters to the model that would account for this behaviour, for example by parametrically modifying the width. We investigate such more complex models and apply those to the KV450 data in \citep{stoelzner/etal:2020}. Instead, we opt to not combine the CZ measurements from the wide-area and deep pencil-beam surveys, as was done before, because such a combination would exactly yield a strongly varying S/N over the peaks for the fourth and fifth tomographic bins \citep[see Fig.~16 from][]{jlvdb/etal:2020}. We therefore use the wide-area surveys exclusively for the first three tomographic bins and the deep pencil-beam surveys for the fourth and fifth bin. This yields relatively symmetric S/N over the redshift range of the peak of each of these bins, and hence fortifies our results against this particular systematic effect. It also allows us to pick different scales over which we evaluate the correlation functions. We use scales of $100\,\mathrm{kpc}<r<1\,\mathrm{Mpc}$ with a full bias correction ($n_{\rm CZ}(z)$, Eq.~\ref{eq:n}), for tomographic bins 1-3, and $30\,\mathrm{kpc}<r<300\,\mathrm{kpc}$ with a correction for the bias of the reference sample only ($\tilde n_{\rm CZ}(z)$, Eq.\ref{eq:ntilde}), for tomographic bins 4~\&~5. These choices are justified by the S/N in the different bins. However, the effect of these choices is captured in our systematic error budget as described in the following.

\begin{figure*}
\centering
\includegraphics[width=0.95\hsize]{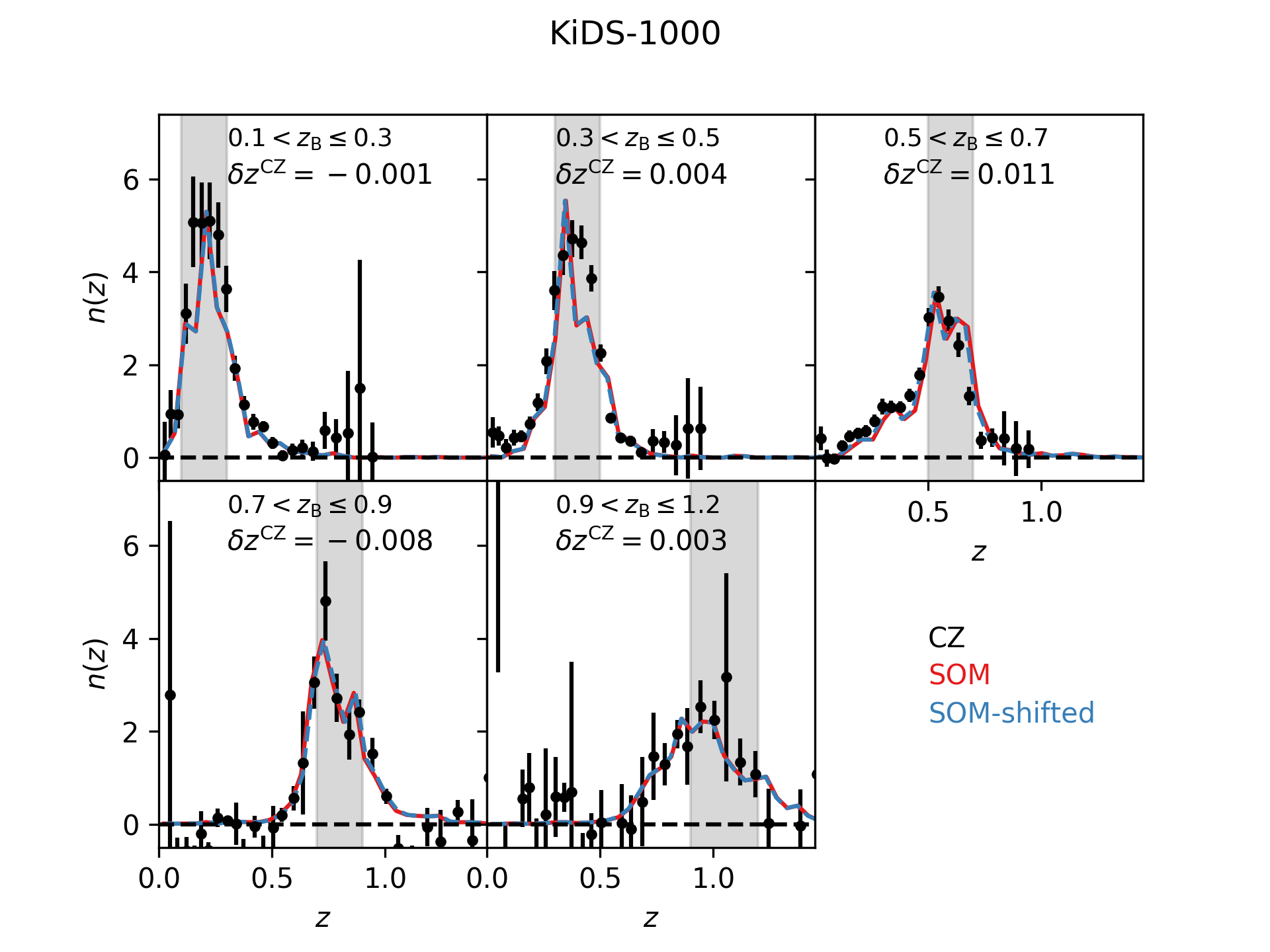}
\caption{\label{fig:res_K1000}Same as Fig.~\ref{fig:res_MICE} but for the KiDS-1000 data.}
\end{figure*}

The model chosen here is quite inflexible. Thus, it cannot be expected to give a good fit (e.g. in terms of reduced $\chi^2$) to the complex CZ data that are affected by residual galaxy bias, variable observing conditions, spectroscopic selection effects, etc., all of which are not modelled. Furthermore, the model itself, being based on the SOM method (Sect.~\ref{sec:SOM}), is noisy, which is not accounted for in our fit. Most importantly, its shape can be slightly different for systematic reasons or due to sample variance. This can be tested on the realistic mocks by using the true redshift distributions as a model. This yields a very good $\chi^2$ arguing that a mismatch in the shape is the most important aspect driving the $\chi^2$ high in realistic situations \citep[see also][for a discussion of the effect of the shape of the model]{jlvdb/etal:2020}. Instead of using only the (possibly unreliable) uncertainties of the best-fit parameters when fitting the SOM $n(z)$ to the CZ data we opt to explore systematic errors by also estimating results for alternative choices of measurement scales, covariance determination, and galaxy bias removal.

Our fiducial approach replicates the methodology of \citet{jlvdb/etal:2020}, with a purely empirical covariance matrix estimated from bootstrap re-sampling, that is approach (A). For the deep pencil-beam surveys (bins 4~\&~5) we also conduct alternative measurements with the simulation-based covariance (C) and a corresponding precision-weighted combination of the different deep fields, as well as measurements at slightly larger scales, $50\,\mathrm{kpc}<r<500\,\mathrm{kpc}$. For the wide-area surveys we consider the covariance alternatives (B) and (C), which allow for a combination of the results from the different surveys via precision-weighting, and alternative measurement scales of $500\,\mathrm{kpc}<r<1.5\,\mathrm{Mpc}$. We also try all different alternatives for bias removal listed in Eqs.~\ref{eq:w}-\ref{eq:n} on the wide fields (bins 1-3) but limit ourselves to the methods referred to in Eqs.~\ref{eq:w}-\ref{eq:ntilde} for the deep fields (bins 4~\&~5). We take the weighted scatter between these alternatives as an estimate of the systematic error inherent to our fiducial choices.

This is far from a perfect estimate of the systematic uncertainty and should not be considered highly precise. Rather it should give a rough idea of possible systematic problems, which is still preferable over quoting a purely statistical uncertainty here. It is clear that this area needs further attention in the future when statistical errors shrink further.

\subsection{Results from MICE mocks}
\label{sec:CZ_res_MICE}
First, we repeat the analysis of \citet{jlvdb/etal:2020} with the idealised reference sample described in Sect.~\ref{sec:MICE}. Using the true redshift distributions as a model yields very small shifts $\delta z^{\rm CZ}_i\la0.005$ for all tomographic bins $i$. This can be regarded as the systematic error floor of our current implementation. We cannot expect the clustering-$z$ with more realistic reference samples and less idealised models to perform any better than this. It should be noted that under these idealised conditions with the very small uncertainties achieved with this dense reference sample, the goodness-of-fit is poor, with values of $\chi^2/\mathrm{d.o.f.}\ga3$ (unlike the case with the more realistic reference samples, where a fit with the true $n(z)$ yields a $\chi^2/\mathrm{d.o.f.}\sim1$, as reported above). We attribute this to the inherent systematic limitations, even with an idealised setup, a general tendency for our errors to be underestimated, and the simplicity of our model. Hence the decision to ignore the goodness-of-fit in the following and estimate the full error budget by exploring alternative analysis choices.

Moving to the fiducial setup, with the first three tomographic bins being calibrated by CZ measurements on the wide-area surveys and the upper two tomographic bins being calibrated by the deep pencil-beam surveys, these numbers vary only very slightly, when the SOM $n(z)$ from one representative line-of-sight (in terms of the mean redshifts of the five tomographic bins) are used as a model. The best-fit solutions and their respective best-fit parameters $\delta z^{\rm CZ}_i$ are reported in Fig.~\ref{fig:res_MICE} and Table~\ref{tab:results} (column 3). Only bin 4 shows a somewhat larger bias of $\delta z^{\rm CZ}_4\sim 0.015$, indicating that the CZ prefers a slightly lower mean redshift than the SOM estimate (in agreement with the value of $\Delta\langle z \rangle^{\rm SOM}$ in that bin). Fitting uncertainties are of the order $\sigma(\delta z^{\rm CZ}_i)\la0.003$, but should not be taken at face value due to the limitations mentioned above.

As described, we explore some alternative scenarios to estimate robust systematic uncertainties for these shifts. The standard deviation between these scenarios ranges from $\sigma_{\rm syst.}(\delta z^{\rm CZ}_i)=0.004$ for bins $i\in(1,2,5)$ to $\sigma_{\rm syst.}(\delta z^{\rm CZ}_4)=0.024$ for bin 4. This indicates that all shifts quoted above are consistent with zero. We report the fitting errors and these systematic error estimates in Table~\ref{tab:results} (column 3).

As we are using the SOM redshift distributions from a single line-of-sight (see Fig.~\ref{fig:SOMnz} for the differences in 100 lines-of-sight), it is clear that there is some residual sample variance that is not fully accounted for in either of the uncertainties quoted in column 3 of Table~\ref{tab:results}. We have, for now, ignored this effect. We do, however, propagate an estimate of the calibration uncertainty due to sample variance in our final CZ results for KiDS-1000 in Sect.~\ref{sec:res_K1000}.  

Figure~\ref{fig:res_MICE} highlights some of the problems encountered with the interpretation of CZ measurements. The uncertainty estimates for the upper two bins are quite noisy due to the small number of deep fields that contribute to the bootstrap re-sampling. Some of these data points clearly influence the fit but the $\delta z^{\rm CZ}$ results suggest that this problem does not lead to an overall large bias. Moreover, the shape of the SOM redshift distributions is somewhat different than the shape suggested by the CZ data points. This mismatch will depend on the line-of-sight chosen for the SOM and highlights the limitations of our modelling. We take the pragmatic stance that, as long as the results for $\delta z^{\rm CZ}$ indicate almost unbiased measurements, these limitations are unimportant for the conclusions drawn in this work.

\subsection{Results from KiDS-1000 data}
\label{sec:res_K1000}
Having verified the methodology from Sect.~\ref{sec:CZ_method} with the mock catalogues in Sect.~\ref{sec:CZ_res_MICE}, we finally apply the clustering-$z$ technique to the KiDS-1000 data. Results for the fiducial setup are reported in Fig.~\ref{fig:res_K1000} and Table~\ref{tab:results}. The best-fit shift parameters $\delta z^{\rm CZ}$ are of the same order as in the simulated analysis, which increases our confidence in the realism of our mock catalogues. There are some subtle differences, such as the bias and systematic scatter in the third bin being slightly larger on the data than on the simulations, with the opposite behaviour in the fourth bin, but the details certainly depend on the line-of-sight chosen for the mocks. Overall the agreement is quite good. We note that there is some mismatch in the shape of the $n(z)$ between the SOM and CZ data for some of the bins. We attribute this partly to sample variance as the SOM $n(z)$ is based on a few lines-of-sight (the deep fields) whereas the clustering-$z$ are estimated from hundreds of square degrees.

We propagate the uncertainty of the mean redshifts of the SOM $n(z)$ into this estimate as we are essentially using a noisy model \citepalias[the noise being a combination of statistical shot noise, cosmological sample variance, and some other contributions; see][]{wright/etal:2020}. We conservatively multiply this SOM uncertainty by a factor of two (column 2 of Table~\ref{tab:results}) to account for limitations in our MICE mocks, in particular the $z<1.4$ redshift limit. Then we add this inflated error and the other errors quoted in column 4 of Table~\ref{tab:results} in quadrature to arrive at the combined uncertainty quoted in the last column.

The magnitude of the uncertainties in the clustering-$z$ measurements is very comparable to the ones from the SOM (compare columns 2 and 4 of Table~\ref{tab:results}). This means that with the KiDS-1000 data set we reach full complementarity between these different approaches of calibrating the $n(z)$.

\begin{figure}
\centering
\includegraphics[width=0.71\hsize]{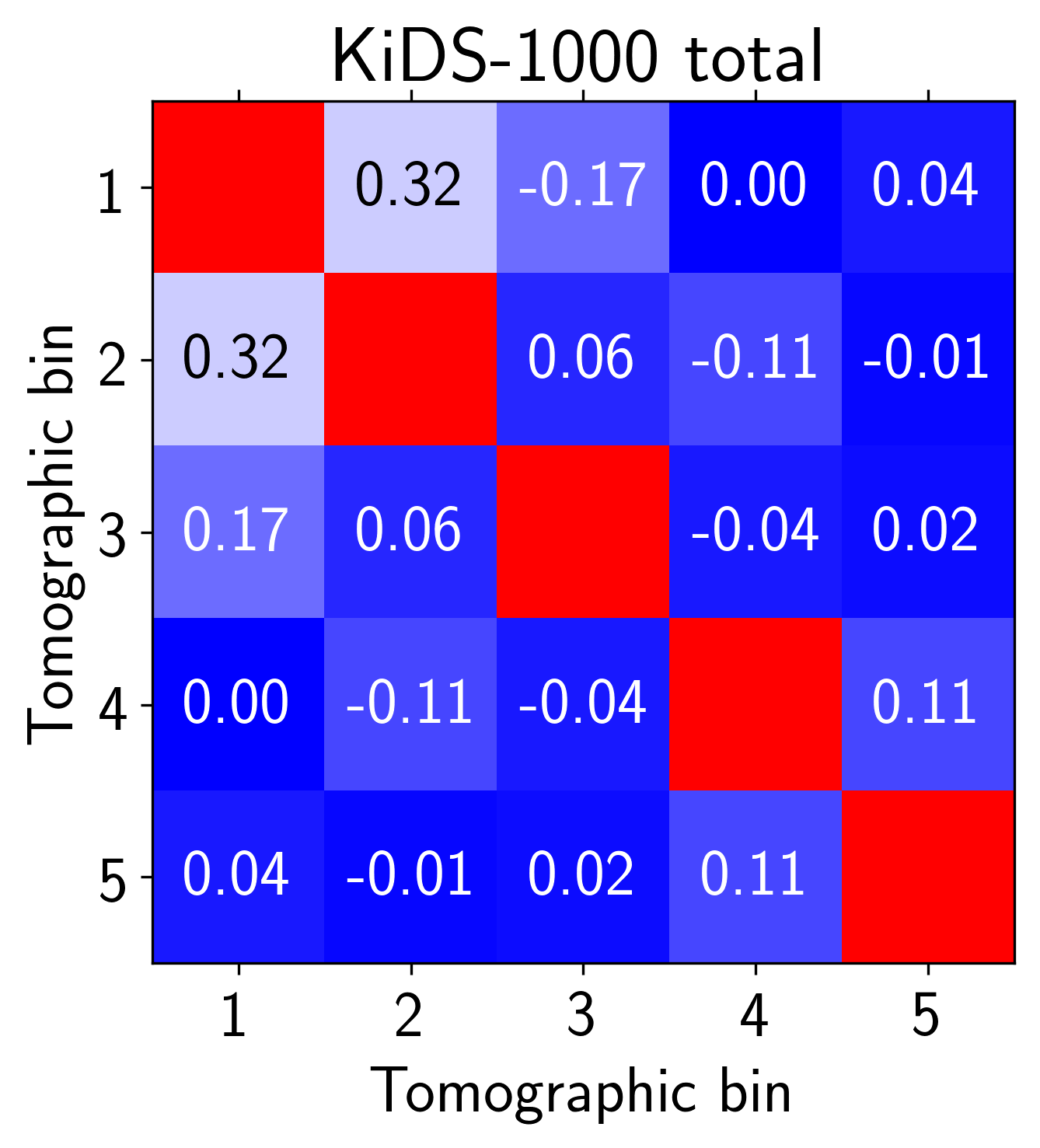}
\caption{\label{fig:K1000_corr}Correlation matrix of the combined uncertainties of the $\delta z^{\rm CZ}_i$ from the CZ analysis of the KiDS-1000 data reported in column 5 of Table~\ref{tab:results}.}
\end{figure}

The uncertainties in the different bins are correlated, with the covariance matrix calculated as the sum of the covariances of the SOM uncertainties, the covariance of the fit parameters $\delta z^{\rm CZ}$, and the covariance of the different alternatives explored in the systematic error estimation.
This combined correlation matrix is shown in Fig.~\ref{fig:K1000_corr}.

\section{Discussion \& summary}
\label{sec:discussion}

The primary calibration method to estimate redshift distributions for the KiDS-1000 cosmology analyses is based on the SOM method. This method projects the high-dimensional magnitude-space into two dimensions so that one can easily identify KiDS galaxies that are not represented by a reference sample. With the same spectroscopic calibration surveys used as a reference, this calibration is almost identical to the one presented in \citetalias{wright/etal:2020}, with the minor exception of updated \emph{lens}fit shape measurement weights and photometric calibration in the current analysis. The accuracy estimates and systematic error discussion of \citetalias{wright/etal:2020} also hold for KiDS-1000 due to the similarity to the KiDS+VIKING-450 data set \citep{wright/etal:2019}. We expect the gold samples defined by the SOM method to be more robustly represented by their corresponding redshift distributions than the full samples used in \citetalias{hildebrandt/etal:2020} were represented by their DIR-estimated $n(z)$. It should be noted that \citet{wright/etal:2020b} showed that cosmological conclusions, in particular the tension w.r.t. Planck, are not strongly affected by this switch in the redshift calibration, while the use of SOM calibration in the KiDS-1000 analyses reduces the redshift calibration systematic uncertainties (compared to the DIR). This represents an important step for systematic error control to keep pace with the growing statistical power of WL surveys.

In previous KiDS analyses, we neglected the correlation of the uncertainties in the mean redshifts of the tomographic bins. Here we report these correlations for the SOM method, as estimated from the covariance of 100 lines-of-sight of the MICE mock catalogues. These correlations will be taken into account in accompanying KiDS cosmological measurements \citep{asgari/etal:2020,heymans/etal:2020,troester/etal:2020} as described in \citet{joachimi/etal:2020}. The SOM $n(z)$ are further validated in \citet{giblin/etal:2020} together with the calibration of the multiplicative shape measurement bias by performing a shear-ratio test \citep{jain/etal:2003,heymans/etal:2012,kitching/etal:2015,schneider:2016} similar to previous KiDS analyses. The $n(z)$ pass this test despite the greater statistical power of KiDS-1000, lending further credence to the stability of the SOM redshift calibration presented here.

Clustering redshifts (CZ) are used as a validation technique for the SOM $n(z)$ in this paper. With unprecedented overlap with spectroscopic surveys over hundreds of square degrees containing more than 300\,000 spectroscopic reference objects, we estimate precise CZ for the KiDS-1000 tomographic redshift bins. The wide-area spec-$z$ reference samples GAMA, BOSS, 2dFLenS, and WiggleZ are used to estimate CZ for the first three tomographic bins with a photo-$z$ range of $0.1<z_\mathrm{B}\le0.7$, whereas the deep pencil-beam surveys zCOSMOS, DEEP2, and VVDS are used for the two high-redshift bins ($0.7<z_\mathrm{B}\le1.2$). This yields a homogeneous S/N of cross-correlation amplitudes as a function of redshift, which is important for the unbiased interpretation of the results.

The same analysis is replicated on mock catalogues based on the MICE simulation to identify and estimate systematic uncertainties (with the caveat that mock galaxies are only available for $z<1.4$). Using the SOM $n(z)$ as a model we fit for residual biases in these primary estimates of the KiDS-1000 redshifts. We find no significant bias in any of the five tomographic bins, neither in the simulated analysis nor on the real KiDS data. The combined uncertainties that are associated with this validation method are at most a factor $\sim2$ larger than the ones estimated for the SOM. As these numbers include the SOM uncertainty (the SOM $n(z)$ are used as a model after all), the CZ method is shown to be fully competitive here.

The most important systematic errors to account for in a CZ analysis are the evolution of the galaxy bias, the non-trivial combination of surveys with different redshift range, density, spatial overlap, and -- connected to this -- model bias from the interpretation of the results with an imperfect model. We mitigate all of these effects and estimate residual systematic uncertainties by analysing a variety of alternative choices for basic analysis parameters (radial measurement scales, covariance estimate, data selection, bias model). These systematic uncertainties are fully propagated into the final CZ results, which constitute an alternative estimate of the KiDS-1000 $n(z)$ to be used in upcoming cosmological measurements.

The work presented here indicates a clear way forward to reach the stringent requirements of stage-IV WL surveys like Euclid \citep{laureijs/etal:2011}, LSST \citep{ivezic/etal:2019}, and RST \citep{spergel/etal:2015}. Given sufficient deep, multi-band photometry, the SOM method allows for a robust gold selection whose accuracy is ultimately only limited by shot noise and competing requirements on the number density of the gold samples. Spectroscopic campaigns like the C3R2 \citep{masters/etal:2017,masters/etal:2019,guglielmo/etal:2020} will push the envelope and allow for increasingly inclusive gold selections with the SOM at further-reduced uncertainties in the redshift distributions.

An interesting addition to these purely spectroscopic approaches to colour-based calibration is offered by the inclusion of high-quality photo-$z$, not as the single calibration source but as a complement to the spectroscopic calibration data already present in the SOM. Multi-wavelength campaigns like the ones in COSMOS \citep{ilbert/etal:2009,ilbert/etal:2013,laigle/etal:2016} can yield exquisite redshift estimates with close to spectroscopic quality but without the drawback of incompleteness.
Even better precision can be obtained from intermediate- or narrow-band surveys such as PAUS \citep{padilla/etal:2019,eriksen/etal:2019} and J-PAS \citep{benitez/etal:2014}, at least at brighter magnitudes. A smart combination of these surveys with the more traditional spectroscopic reference samples in a colour-based calibration like the SOM will mitigate the individual weaknesses of these catalogues and leverage their complementary advantages.

The future of the CZ technique looks similarly bright. Most of the limiting systematic effects seem to be understood by now and mitigation techniques have been established. The interpretation with a suitable model and subsequent estimation of realistic uncertainties is currently the biggest methodological problem to overcome. On the data side, the redshift range covered by wide-area surveys is still not sufficient to leverage the full potential of CZ. Currently, only the cores of the redshift distributions of typical weak lensing source samples can be calibrated with CZ. But with the advent of new spectroscopic facilities like DESI\footnote{Dark Energy Spectroscopic Instrument; \url{www.desi.lbl.gov}} \citep{desi16}, 4MOST\footnote{\url{www.4most.eu}} \citep{richard19}, WEAVE\footnote{\url{www.ing.iac.es/weave/}} \citep{dalton:2016}, and PFS\footnote{Subaru Prime Focus Spectrograph; \url{https://pfs.ipmu.jp}} \citep{takada14} this situation will improve and the crucial calibration of high-redshift tails will become possible at high precision.

All of these data-related efforts need to be accompanied by improved mock catalogues and better theoretical understanding. In terms of mocks, larger volumes, higher redshifts, even more realistic galaxy colours, and a realistic integration of galaxy colours and shapes is needed. On the theoretical side, the standard practice in the analysis of weak lensing surveys regards the work presented here as calibration steps that are carried out before the main cosmological inference. In the future, this clear distinction could be broken up, with parts or all of this calibration being integrated into the inference pipeline itself \citep{bernstein:2009}. This is more obvious for CZ, which represents ``just another two-point function'' to model and fit, but such an integration can also be imagined for the colour-based calibration approach. While systematic error control is an issue in such integrated approaches, the optimal use of information in the data for example through Bayesian-hierarchical modelling \citep{sanchez/etal:2019,alarcon/etal:2019} makes this idea extremely attractive for established methods that have left the exploratory stage.

The work presented here means that the KiDS-1000 cosmological analyses based on these weak lensing source samples will not be limited in their statistical power by the uncertainties in the redshift distributions. The constant progress and the developments sketched above make it seem realistic to meet the extremely tight requirements on the redshift calibration for stage-IV surveys a few years from now, a situation that seemed almost inconceivable not too long ago.

\begin{acknowledgements}

We are grateful to the anonymous referee for some suggestions that improved the paper. \\
  
We are indebted to the staff at ESO-Garching and ESO-Paranal for managing the observations at VST and VISTA that yielded the data presented here. Based on observations made with ESO Telescopes at the La Silla Paranal Observatory under programme IDs 177.A-3016, 177.A-3017, 177.A-3018, 179.A-2004, 298.A-5015, and on data products produced by the KiDS consortium. \\

The 2dFLenS survey is based on data acquired through the Australian Astronomical Observatory, under program A/2014B/008. It would not have been possible without the dedicated work of the staff of the AAO in the development and support of the 2dF-AAOmega system, and the running of the AAT. \\

GAMA is a joint European-Australasian project based around a spectroscopic campaign using the Anglo-Australian Telescope. GAMA is funded by the STFC (UK), the ARC (Australia), the AAO, and the participating institutions. The GAMA website is \url{http://www.gama-survey.org/}.\\

Funding for the Sloan Digital Sky Survey IV has been provided by the Alfred P. Sloan Foundation, the U.S. Department of Energy Office of Science, and the Participating Institutions. SDSS-IV acknowledges
support and resources from the Center for High-Performance Computing at
the University of Utah. The SDSS web site is \url{www.sdss.org}.\\

We are grateful to the zCOSMOS team to give us early access to additional deep spec-$z$ that were not available in the public domain. zCOSMOS is based on observations made with ESO Telescopes at the La Silla or Paranal Observatories under programme ID 175.A-0839. This research has made use of the zCOSMOS database, operated at CeSAM/LAM, Marseille, France.\\

Funding for the DEEP2 Galaxy Redshift Survey has been
provided by NSF grants AST-95-09298, AST-0071048, AST-0507428, and AST-0507483 as well as NASA LTSA grant NNG04GC89G.\\

This research uses data from the VIMOS VLT Deep Survey, obtained from the VVDS database operated by Cesam, Laboratoire d'Astrophysique de Marseille, France.\\

We acknowledge support from European Research Council grants 770935 (HH, JLvdB, AHW, AD) and 647112 (CH, MA, BG, CL, TT), the European Union's Horizon 2020 research and innovation programme under the Marie Sk{l}odowska-Curie grant agreement No 797794 (TT), as well as the Deutsche Forschungsgemeinschaft (HH, Heisenberg grant Hi 1495/5-1).
MB is supported by the Polish Ministry of Science and Higher Education through grant DIR/WK/2018/12, and by the Polish National Science Center through grant no. 2018/30/E/ST9/00698.
CH acknowledges support from the Max Planck Society and the Alexander von Humboldt Foundation in the framework of the Max Planck-Humboldt Research Award endowed by the Federal Ministry of Education and Research.
KK acknowledges support by the Alexander von Humboldt Foundation.
HYS acknowledges the support from NSFC of China under grant 11973070, the Shanghai Committee of Science and Technology grant No.19ZR1466600 and Key Research Program of Frontier Sciences, CAS, Grant No. ZDBS-LY-7013.
JTAdJ is supported by the Netherlands Organisation for Scientific Research (NWO) through grant 621.016.402.

\\
\textit{Author Contributions:} All authors contributed to the development and writing of this paper. The authorship list is given in three groups: the lead authors (HH, JLvdB, AHW), followed by two alphabetical groups. The first alphabetical group includes those who are key contributors to both the scientific analysis and the data products. The second group covers those who have either made a significant contribution to the data products or to the scientific analysis.
\end{acknowledgements}

\bibliographystyle{aa}

\bibliography{K1000-photoz}

\appendix

\section{Toy model for data with variable S/N}
\label{sec:toy_model}
Here we illustrate with a simple toy model how variable S/N can bias a model fit. This situation is quite common in clustering-$z$ measurements that typically exhibit a large number of reference galaxies at low redshift and a small number at high redshift due to the difficulties of measuring redshifts for high-$z$ galaxies.

\begin{figure}
 \centering
 \includegraphics[width=\hsize,clip=true,trim=0cm 0cm 38.7cm 0cm]{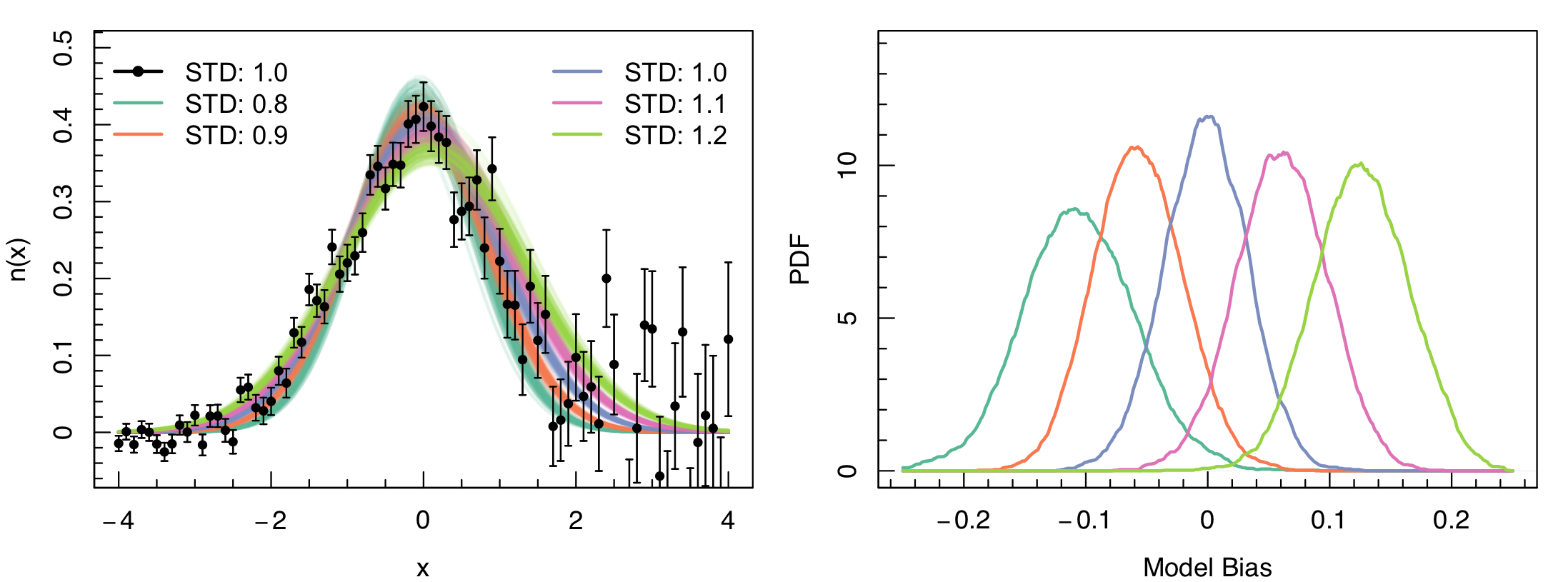}
 \includegraphics[width=\hsize,clip=true,trim=38.7cm 0cm 0cm 0cm]{figures/cc_toy_model}
 \caption{\label{fig:toy_model}Toy model to illustrate the effect of variable S/N on model fits. \emph{Top}: Black data points correspond to one noise realisation with decreasing S/N. The blue lines correspond to fits with a model of perfect width whereas the teal and orange lines correspond to models that are to narrow and the magenta and green lines correspond to models that are too wide. \emph{Bottom}: If the model has the correct width the mean of the best fit is on average unbiased (blue) whereas it is on average biased low if the model is too narrow (teal and orange) and biased high if the model is too broad (magenta and green).}
\end{figure}

In Fig.~\ref{fig:toy_model} we show a simulated data set that is based on a normal distribution with the S/N decreasing as a function of $x$. Fitting a model with a shift parameter and a free amplitude to different noise realisations yields the coloured lines. If the model has the correct width (i.e. standard deviation $\mathrm{STD}=1$) the model fits (blue lines) are on average unbiased as shown in the bottom panel. If the model is too narrow ($\mathrm{STD}<1$) the model fits (teal and orange) are biased low whereas if the model is too broad ($\mathrm{STD}>1$) the model fits (magenta and green) are biased high.

This observation led to our decision to analyse the clustering-$z$ of the wide and deep fields separately. The number of reference galaxies in the two sets is just too different so that a sharp drop in S/N is observed at the transition redshift ($z\sim0.8$). Analysing the wide and deep fields together and using the SOM (DIR) $n(z)$, whose widths are typically to small (large), would result in a similar model bias as shown in Fig.~\ref{fig:toy_model}.

One alternative would be to randomly subsample the Wide data to homogenise the S/N. We leave this idea to future work.

\end{document}